\documentclass[acmtog,authorversion]{acmart}

\usepackage{booktabs} 

\usepackage{bm}
\usepackage{amsmath}
\usepackage{amsthm}
\usepackage{graphicx}
\usepackage{natbib}
\usepackage{enumerate}
\usepackage{float}
\usepackage{subfig}

\usepackage{multirow}
\usepackage{rotating}

\usepackage{setspace}
\usepackage{tikz}
\usepackage{pgf}
\usepackage{tkz-graph}
\usetikzlibrary{arrows}
\usetikzlibrary{shapes}
\usepackage{verbatim}

\newtheorem{myprop}{Proposition}

\newtheorem{mythm}{Theorem}

\theoremstyle{definition}
\theoremstyle{remark}

\newtheorem{assumption}{Assumption}
\theoremstyle{definition}
\newtheorem{fact}{Fact}
\theoremstyle{definition}
\newtheorem{mydef}{Definition}

\usepackage[ruled]{algorithm2e} 

\SetAlFnt{\small}
\SetAlCapFnt{\small}
\SetAlCapNameFnt{\small}
\SetAlCapHSkip{0pt}
\IncMargin{-\parindent}

\newcommand{\prefprofile}{\mathbf{u}}

\newcommand{\benmatrix}{benefits matrix}
\newcommand{\B}{\mathbf{B}}

\DeclareMathOperator{\trace}{trace}

\newcommand{\be}{\begin{equation}}
\newcommand{\ee}{\end{equation}}
\newcommand{\bes}{\begin{equation*}}
\newcommand{\ees}{\end{equation*}}
\newcommand{\mbf}{\mathbf}

\newcommand{\R}{\mathbb R}

\definecolor{Teal}{RGB}{0,128,128}
\usepackage{hyperref}
\hypersetup{colorlinks,breaklinks,
	urlcolor=Teal,
	linkcolor=Teal,
	citecolor=Teal,
	}

\setcopyright{usgovmixed}

\acmDOI{0000001.0000001_2}

\begin{document}


\title{{\fontsize{17pt}{15pt}\selectfont A Network Approach to Public Goods \hfill \fontsize{14pt}{16pt}\selectfont  \sc{Matthew Elliott and Benjamin Golub$^{\dagger}$}} \\  \vspace{-.05in} {\fontsize{15pt}{18pt}
			\selectfont \sc{A short summary} \hfill \href{http://bengolub.net/wp-content/uploads/2020/05/jmp.pdf}{[Full paper]}  }}

\renewcommand\shortauthors{Elliott and Golub}

\maketitle



\renewcommand{\thefootnote}{\fnsymbol{footnote}}
\footnotetext[2]{\emph{Date}: July 2021. Elliott: Cambridge. Golub: Northwestern.}
\renewcommand{\thefootnote}{\arabic{footnote}}

Consider an economy in which agents ($N = \{1,2,\ldots,n\}$) each choose a level of effort, $a_i \geq 0$, toward an activity, such as mitigating pollution, that generates externalities. Agent $i$ has a utility function $u_i : \R_{+}^n \to \R$ (where $u_i$ is concave and continuously differentiable) mapping everyone's efforts to payoffs.

The incidence of externalities may be heterogeneous across those affected. How do the asymmetries and heterogeneities shape outcomes? We focus on a particular class of outcomes---Pareto efficient outcomes---which may be reached, e.g., by efficient multilateral negotiations.  Our main thesis is that we can gain insight on the welfare economics of this setting by studying a network reflecting externalities. For any action profile $\mathbf{a}$, we construct a network in which the agents are nodes and the weighted links among them record the marginal externalities of actions. In particular, the link or arrow from agent $i$ to $j$ reflects how much marginal benefit $i$ can confer on $j$ by increasing $i$'s action. Our main results relate important network-theoretic statistics of the benefits network to economic properties of the environment. We show that the size of the largest eigenvalue of the benefits network diagnoses the potential for Pareto improvements. The Pareto weights at efficient outcomes turn out to be equal to agents' eigenvector centralities in the network. Moreover, in an important class of negotiated outcomes, agents' efficient \emph{actions} are also proportional to their eigenvector centralities in the transpose of the network.

\section{Main Assumptions and Definitions}\label{sec:assumptions}

The following two assumption capture that agents can, at a cost, provide nonrival positive externalities to each other.

\begin{assumption}[Costly Actions]\label{as:ca}
	Each agent finds it costly to invest  effort, holding others' actions fixed: $\frac{\partial u_i}{\partial a_i}(\mathbf{a})< 0 $  for any $\mathbf{a} \in \R_{+}^n$ and $i \in N$.
\end{assumption}

\begin{assumption}[Positive Externalities]\label{as:pe}
	Increasing any agent's action level weakly benefits all other agents:  $\frac{\partial u_i}{\partial a_j}(\mathbf{a})\geq 0 $  for any $\mathbf{a} \in \R_{+}^n$ whenever $j \neq i$.
\end{assumption}

Imagine for a moment that agents choose actions noncooperatively, playing a Nash equilibrium in the environment described. Costly Actions implies that the unique Nash equilibrium has $a_i=0$ for each $i$. Except in the trivial case where the action profile $\bm{0}$ is already Pareto efficient, this means that there are Pareto gains from increasing actions together through some kind of favor-trading that improves on the static Nash outcome.\footnote{One interpretation of the action profile $\mathbf{a}=\bm{0}$ is as a status quo at which negotiations begin.  An alternative interpretation is that it is a Nash equilibrium in which everyone has already exhausted their private gains from exerting effort.}

The Jacobian, $\mathbf{J}(\mathbf{a})$, is the $n$-by-$n$ matrix whose $(i,j)$ entry is $J_{ij}(\mathbf{a})= \partial u_i(\mathbf{a}) / \partial a_j$. The \emph{\benmatrix}\ $\B(\mathbf{a;\mathbf{u}})$ is then defined as follows:
$$B_{ij}(\mathbf{a};\mathbf{u}) =
\begin{cases}
\frac{J_{ij}(\mathbf{a;\mathbf{u}})}{-J_{ii}(\mathbf{a;\mathbf{u}})}  & \text{if $i\ne j$}\\
0 & \text{otherwise}.
\end{cases} $$

When $i \neq j$, the quantity $B_{ij}(\mathbf{a;\mathbf{u}})$ is $i$'s marginal rate of substitution between decreasing own effort and receiving help from $j$. In other words, it is how much $i$ values the help of $j$, measured in the number of units of effort that $i$ would be willing to put forth in order to receive one unit of $j$'s effort.

Since $J_{ii}(\mathbf{a};\mathbf{u})<0$ by Costly Actions, the benefits matrix is well-defined. Costly Actions and Positive Externalities, assumptions which we will maintain throughout, imply that it is entrywise nonnegative. We also assume throughout that $\mathbf{B}(\mathbf{a})$ is irreducible, meaning that it is not possible to find an outcome and partition society into two nonempty groups such that, at that outcome, one group does not derive any marginal benefit from the effort of the other group.

\section{Efficiency and the Spectral Radius} \label{sec:efficiency}

This section shows that an important statistic of the benefits network, the size of the largest eigenvalue, can be used to diagnose whether an outcome is Pareto efficient. For any nonnegative matrix $\mathbf{M}$, we define $r(\mathbf{M})$ as the maximum of the magnitudes of the eigenvalues of  $\mathbf{M}$, also called the \emph{spectral radius}. This quantity can be interpreted as a single measure of how expansive a matrix is as a linear operator---how much it can scale up vectors that it acts on. 

We also recall the definition of eigenvector centrality, an important notion for us.\footnote{We freely use the Perron--Frobenius theorem here and throughout.} For any nonnegative, irreducible matrix $\mathbf{M}$, left-hand \emph{eigenvector centralities} are entries of a positive vector $\bm{\ell}$ such that $\bm{\ell} \mathbf{M} = \lambda \bm{\ell}$ for a constant $\lambda>0$ (which in fact must be equal to $\rho(\mathbf{M})$). Such an $\bm{\ell}$ turns out to be unique, up to multiplication by a positive scalar. Analogously, right-hand \emph{eigenvector centralities} are entries of a positive vector $\mathbf{r}$ such that $\mathbf{M}\mathbf{r} = \lambda \mathbf{r}$. Eigenvector centrality captures the idea that central nodes are those that are adjacent to other central nodes. More precisely, $i$'s centrality is a weighted sum of $i$'s link weights, weighted by the centralities of the others involved in those links. For $\bm{\ell}$, the link weights are the numbers in \emph{column} $i$ of $\mathbf{M}$, while for $\mathbf{r}$, the link weights are the numbers in \emph{row} $i$ of $\mathbf{M}$. 

\begin{myprop} \label{prop:pareto}
	$ $
	\begin{enumerate}
		\item[(i)] An interior action profile $\mathbf{a} \in \R_{++}^n$ is Pareto efficient  if and only if the spectral radius of $\B(\mathbf{a})$ is $1$. At such an $\mathbf{a}$, the  eigenvector centralities $\bm{\theta}$ defined by the property  $\bm{\theta} \mathbf{B}(\mathbf{a}) = \bm{\theta}$ are the Pareto weights such that $\mathbf{a}$ maximizes $\sum_i \theta_i u_i(\mathbf{a})$.  
		\item[(ii)] The outcome $\mathbf{0}$ is Pareto efficient if and only if  $r(\B(\mathbf{0})) \leq 1$.
	\end{enumerate}
\end{myprop}

We will discuss the proof of (i) only. The idea of the argument is that if the spectral radius is greater than $1$, a Pareto improvement can be constructed in which one agent increases his action, generating benefits for others; then others ``pass forward'' some of the benefits they receive by increasing their own actions.  

Fix any $\mathbf{a} \in \R_{++}^n$ (which we will often suppress as an argument) and let  $\rho$ denote the spectral radius of $\mathbf{B}(\mathbf{a})$. Then by the Perron--Frobenius theorem and the maintained assumptions, there is a $\mathbf{d}\in \mathbb{R}^n_{++}$ such that $\mathbf{B}\mathbf{d}=\rho \mathbf{d}$. Multiplying each row of this system by $-J_{ii}(\mathbf{a})$,
$$  \sum_{j \neq i} \tfrac{\partial u_i}{\partial a_j} d_j + \rho \tfrac{\partial u_i}{\partial a_i} d_i = 0 \quad \forall i.$$
If $\rho>1$, then using the assumption of Costly Actions ($\frac{\partial u_i}{\partial a_i}<0$) we deduce
\begin{equation}   \sum_{j \neq i} \tfrac{\partial u_i}{\partial a_j} d_j + \tfrac{\partial u_i}{\partial a_i} d_i > 0 \quad \forall i,\label{pareto-inequality}\end{equation}
showing that a slight change where each $i$ increases his action by the amount $d_i$ yields a Pareto improvement. The vector $\mathbf{d}$ describes the relative magnitudes of contributions to make the passing forward of benefits work out to achieve a Pareto improvement. Note that it is key to the argument that $\mathbf{d}$ is positive. The conditions of the Perron--Frobenius theorem guarantee the positivity of $\mathbf{d}$.  If $\rho<1$, we reason similarly to conclude the inequality (\ref{pareto-inequality}) when we set $\mathbf{d}$ to be \emph{minus} the Perron vector of $\mathbf{B}$---i.e., when each $i$ slightly \emph{decreases} his action by the amount $|d_i|$. Thus Pareto efficiency implies $\rho(\mathbf{B}(\mathbf{a}))=1$.

Conversely, we now show that if $\rho(\mathbf{B}(\mathbf{a}))=1$ then $\mathbf{a}$ is Pareto efficient. By the Perron--Frobenius theorem, if $\rho(\mathbf{B}(\mathbf{a}))=1$ there is a \emph{left}-hand eigenvector $\bm{\theta}$ of $\mathbf{B}(\mathbf{a})$, with all positive entries, satisfying  $\bm{\theta}\mathbf{B}(\mathbf{a}) = \bm{\theta}$. This can be rearranged into the equation $\bm{\theta} \mathbf{J}(\mathbf{a})=\mathbf{0}$, which is the system of first-order conditions for choosing $\mbf{a}$ to maximize $ \sum_i \theta_i u_i(\mathbf{a})$. Since the first-order conditions hold for the vector of weights $\bm{\theta}$ and the maximization problem is concave, it follows that $\mathbf{a}$ is Pareto efficient. 

The above argument also shows that whenever $\mathbf{a}$ is efficient, the vector $\bm{\theta}$ of left-hand eigenvector centralities of $\mathbf{B}(\mathbf{a})$ is such that $\mathbf{a}$ maximizes $ \sum_i \theta_i u_i(\mathbf{a})$. Intuitively,  $\theta_i = \sum_j \theta_j B_{ji}$ says $i$'s weight (proportional to the planner's disutility of his cost) equals the total benefits he can confer on others, weighted by their $\theta_j$; if this were not so, the planner would want $i$ to work harder.

\subsection{Application: Essential agents} \label{sec:essential}

Are there any agents that are essential to negotiations in our setting and, if so, how can we identify them? The efficiency result above suggests a simple approach to this question. Suppose for a moment that a given agent may be exogenously unable to take any action other than $a_i=0$. How much does such an exclusion hurt the prospects for voluntary cooperation by the other agents?

Without agent $i$, the benefits matrix at the status quo of $\mathbf{0}$ is equal to the original $\mathbf{B}(\mathbf{0})$ with each entry in that row in $i$'s column set to $0$. Call that matrix $\mathbf{B}^{[-i]}(\mathbf{0})$. Its spectral radius is no greater than that of  $\mathbf{B}(\mathbf{0}).$ In terms of consequences for efficiency, the most dramatic case is one in which the spectral radius of $\mathbf{B}(\mathbf{0})$ exceeds $1$ but the spectral radius of  $\mathbf{B}^{[-i]}(\mathbf{0})$ is less than $1$. Then by Proposition \ref{prop:pareto}(ii), a Pareto improvement on $\mathbf{0}$ exists when $i$ is present but not when $i$ is absent.

Thus \emph{agent $i$'s participation is essential to achieving any Pareto improvement on the status quo precisely when his removal changes the spectral radius of the benefits matrix at the status quo from being greater than $1$ to being less than $1$}. To illustrate this, consider the following example in which $N=\{1,2,3,4\}$.

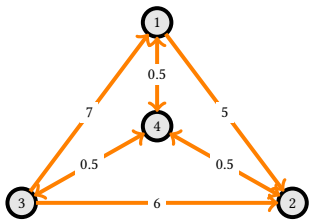
\begin{figure}[ht]
	
	\centering
		\begin{tikzpicture}[baseline={([yshift=-.5ex]current bounding box.center)},scale=0.6, every node/.style={transform shape}]
		\SetVertexNormal[Shape      = circle,
		FillColor = gray!20,
		LineWidth  = 1.5pt]
		\SetUpEdge[lw         = 1.5pt,
		color      = orange,
		labelcolor = white]
		
		\tikzset{node distance = 1.6in}
		
		\tikzset{VertexStyle/.append  style={fill}}
		\Vertex[x=0,y=0]{1}
		\Vertex[x=3,y=-4]{2}
		\Vertex[x=-3,y=-4]{3}
		\Vertex[x=0,y=-2.3]{4}
		\tikzset{EdgeStyle/.style={->}}
		\Edge[label=5](1)(2)
		\Edge[label=6](3)(2)
		\Edge[label=7](3)(1)
		\tikzset{EdgeStyle/.style={<->}}
		\Edge[label=0.5](1)(4)
		\Edge[label=0.5](4)(3)
		\Edge[label=0.5](4)(2)
		\end{tikzpicture} 
	
	\caption{A graph corresponding to a benefits matrix in which agent \#4 is essential despite providing smaller benefits than the others.}
	\label{fig:three-cycle-2}
\end{figure}

The import of the example is that agent $4$, even though he confers the smallest marginal benefits, is the only essential agent. Without him, there are no cycles at all and the spectral radius of the corresponding benefits matrix $\mathbf{B}^{[-4]}(\mathbf{0}) $ is $0$. On the other hand, when 4 is present but any one other agent ($i \neq 4$) is absent, then there is a cycle whose edges multiply to more than $1$, and the spectral radius of $\mathbf{B}^{[-i]}(\mathbf{0})$ exceeds $1$.  Thus, the participation of a seemingly ``small'' agent in negotiations can make an essential difference to the ability to improve on the status quo when that agent completes cycles in the benefits network.

\subsection{Spectral Radius in Terms of Cycles}

 A standard fact permits a general expression of the role cycles play in the spectral radius (for background and a proof, see, e.g., \cite{milnor}).

\begin{fact}\label{fact:cycles}\label{fact:monotonic}
For any nonnegative matrix $\mathbf{M}$, its spectral radius $r(\mathbf{M})$ is equal to $ \limsup_{\ell \to \infty} \trace \left(\mathbf{M}^\ell\right)^{1/\ell} .$

\end{fact}

For a directed, unweighted adjacency matrix $\mathbf{M}$, the quantity $\trace \left(\mathbf{M}^\ell\right)$ counts the number of cycles of length $\ell$ in the corresponding network. More generally, for an arbitrary matrix $\mathbf{M}$ it measures the strength of all cycles of length $\ell$ by taking the product of the edge weights for each such cycle, and then summing these values over all such cycles. Thus, by Fact \ref{fact:cycles}, the (suitably adjusted) total value of long cycles provides an asymptotically exact estimate of the spectral radius.

The essential agents discussed in the last section will be those that are present in sufficiently many of the high value cycles in the network. Relatedly, a single weak link in a cycle will dramatically reduce the value of that cycle. Thus networks with an imbalanced structure, in which it is rare for those agents who could confer large marginal benefits on others to be the beneficiaries of others' efforts, will have a lower spectral radius and there will be less scope for cooperation.

\subsection{Application: Subdividing Negotiations} \label{sec:subdividing}
Consider an arbitrary Pareto efficient outcome $\mathbf{a}^*$ that a planner would like to achieve. Suppose that  the agents cannot negotiate in the full group (perhaps because a large negotiation is too costly) and are divided into two subsets, $M$ and $M^c$;  $\mathbf{a}^*$ is proposed to each. Then each group can contemplate  deviations from  $\mathbf{a}^*$ that are Pareto-improving \emph{for that group}. How cheaply can a planner incentivize agents to stay with the original outcome rather than deviate? 

For simplicity, set $J_{ii}(\mathbf{a})=-1$ for each $i$ and all $\mathbf{a}$, and assume, for this subsection only, that the planner may use transfers of a numeraire that enters each agent's utility additiviely. New payoffs are $ \widetilde{u}_i(\mathbf{a})=u_i(\mathbf{a})+m_i(\mathbf{a}),$ where $m_i(\mathbf{a})$ must be nonnegative. The profile $(m_i)_{i \in N}$ \emph{deters deviations from $\mathbf{a}^*$} if the restriction of $\mathbf{a}^*$ to $M$ is Pareto efficient for the population $M$ with preferences $(\widetilde{u}_i(\mathbf{a}))_{i \in N}$, and if the analogous statement holds for $M^c$. We care about bounding the \emph{cost of separation} $c_M(\mathbf{a}^*)$, defined as the infimum of $ \sum_{i \in N} m_i(\mathbf{a}^*)$---payments made by the planner at the implemented outcome---taken over all profiles $(m_i)_{i \in N}$ that deter deviations from $\mathbf{a}^*$. 
\begin{proposition}\label{prop:splitting} Consider a Pareto efficient outcome $\mathbf{a}^*$, and let $\bm{\theta}$ be the corresponding Pareto weights. Then$$c_M(\mathbf{a}^*)\leq  \sum\frac{\theta_i}{\theta_j} B_{ij}(\mathbf{a}^{*})a_j^*,$$ where the summation is taken over all ordered pairs $(i,j)$ such that one element is in $M$ and the other is in $M^c$. \end{proposition}

Holding $\mathbf{a}^*$ fixed, the bound in the proposition is small when the network given by $\mathbf{B}(\mathbf{a}^*)$ has small total weight on links across groups---i.e. when the \emph{cut} between them is small. Note that it is the properties of \emph{marginal} benefits that are key. The propositions shows that a negotiation can be very efficiently separable even when the separated groups provide large total (i.e., inframarginal) benefits to each other. The question of when one can find a split with this property is discussed in a large literature in applied mathematics on spectral clustering and the spectral gap. One conclusion is that if there is an eigenvalue of $\mathbf{B}(\mathbf{a}^*)$  near its largest eigenvalue ($1$ in this case, since $\mathbf{a}^*$ is efficient) then such a split exists \citep{MeyerUncoupling}.

\section{Linear Favor-Trading: Lindahl Outcomes} \label{sec:lindahl}\label{sec:characterization}

In this section, we focus on a particular class of Pareto efficient solutions. The idea is that a public good would be provided efficiently if we could replicate markets, allowing agents to trade personalized externalities at market prices. Then  trade should occur up to the point where the marginal social benefits of agents' efforts equal the marginal cost.  Following this idea, we price actions and look for a Walrasian equilibrium. \citet{penta} provides a way to implement such outcomes via a bargaining game. In the Online Appendix of the full paper, we discuss bargaining and implementation theory foundations for this solution in detail.

To define Lindahl outcomes, let $\mathbf{P}$ be an $n$-by-$n$ matrix of prices, with $P_{ij}$  (for $i \neq j$)\ being the personalized price $i$ pays to $j$ per unit of $j$'s effort. Let $Q_{ij}$ be how much $i$ purchases of $j$'s effort at this price. The total expenditure of $i$ on other agents' efforts is $ \sum_j P_{ij}Q_{ij}$ and the total income that $i$ receives from other agents is $ \sum_j P_{ji} Q_{ji}$. Market-clearing requires that all agents $i\ne j$ demand exactly the same effort from agent $j$, and so $Q_{ij}=a_j$ for all $i$ and all $j\ne i$. Incorporating these market clearing conditions, agent $i$ faces the budget constraint
\be   \sum_{j:j\neq i} P_{ij}a_j \leq  a_i   \sum_{j : j \neq i} P_{ji}  \label{eq:budget-balance} \tag{$\text{BB}_i(\mathbf{P})$}.\ee

The Lindahl solution requires that, at prevailing prices, the outcome is each agent's most preferred among those satisfying this budget balance condition:

\begin{mydef}\label{def:lindahl}
	An action profile $\mathbf{a}^*$ is a \emph{Lindahl outcome} for a preference profile $\prefprofile$ if there are prices $\mathbf{P}$ so that the following conditions hold for every $i$:
	\begin{enumerate}
		\item[(i)] \ref{eq:budget-balance} is satisfied when $\mathbf{a}=\mathbf{a}^*$;
		\item[(ii)] for any $\mathbf{a}$ such that the inequality $\text{BB}_i(\mathbf{P})$ is satisfied, we have $u_i(\mathbf{a}^*)\geq u_i(\mathbf{a})$.
	\end{enumerate}
\end{mydef}

The main result in this section,  Theorem \ref{thm:main}, relates agents' contributions in Lindahl outcomes to how central they are in the network of externalities.

\begin{mydef}
	An action profile $\mathbf{a} \in \R_{+}^n$ has the \emph{centrality property} (or is a \emph{centrality action profile}) if $\mbf{a} \neq \mbf{0}$ and $\B(\mathbf{a}) \mathbf{a}  =  \mathbf{a}$.
\end{mydef}

This condition holds if $\mathbf{a}$ is a right-hand eigenvector of $\mathbf{B}(\mathbf{a})$ with eigenvalue $1$, i.e., the players' actions are their right-hand eigenvector centralities in $\mathbf{B}(\mathbf{a})$. The centrality property says that, for each $i$, we have
\begin{equation} a_i =  \textstyle \sum_{j\in N} B_{ij} a_j. \label{eq:centrality} \end{equation}
Equation (\ref{eq:centrality}) asserts that  each agent's contribution is a weighted sum of the other agents' contributions, where the weight on $a_j$ is proportional to the marginal benefits that $j$ confers on $i$. 

\begin{mythm} \label{thm:main} \label{thm:main-theorem}
	The following are equivalent for $\mathbf{a} \in \R_{+}^n \setminus \{\bm{0}\}$: 
	\begin{enumerate}
		\item[(i)] $\B(\mathbf{a}) \mathbf{a}  =  \mathbf{a}$, i.e., $\mathbf{a}$ has the centrality property;
		\item[(ii)] $\mathbf{a}$ is a Lindahl outcome.
	\end{enumerate}
\end{mythm}

Informally, Lindahl outcomes exist because they are Walrasian equilibria, and Walrasian equilibria exist for the usual reasons.\footnote{In fact, in our setting standard proofs do not go through because of their boundedness requirements, but in \citet{elliott2019network} we establish the existence of Lindahl outcomes generally, using our main characterization.}

\subsection{Coalitional Deviations: A Core Property}\label{sec:coalitional_deviations}

As we are modeling negotiations, we can ask whether some subset of the agents could do better by breaking away and coming to some other agreement among themselves. Suppose that conditional on a deviation, the non-deviating agents respond with their unilateral best-response actions of $0$. Outcomes where no coalition has a deviation that leaves all its members better off (assuming this response) are said to be in the \emph{core}. 

We show that if $\mathbf{a} \in \R_+^n$ has the centrality property, then it is in the core. The reasoning closely follows that of \citet{shapley1969core}, who show an equivalence between the core we have just defined the standard core of the above-described artificial economy with personalized externalities. The core property of Walrasian economies then completes the proof.

Thus, Lindahl outcomes are appealing because they are guaranteed to be in the core; this is intuitive, since by Theorem 1, both individuals and groups contribute to the extent that they value contributions  made by others.


\subsection{An Outline of the Proof of Theorem \ref{thm:main} } \label{sec:main-intuition}

The key fact for the more difficult ``if'' part is that the system of equations $\mathbf{B}(\mathbf{a}^*) \mathbf{a}^*  = \mathbf{a}^*$  allows us to extract Pareto weights that support the outcome $\mathbf{a}^*$ as efficient; using those Pareto weights and the Jacobian, we can construct prices that support $\mathbf{a}^*$ as a Lindahl outcome.

Now in more detail:
Suppose we have a nonzero $\mathbf{a}^*$ so that $\mathbf{B}(\mathbf{a}^*) \mathbf{a}^*  = \mathbf{a}^*$. The Perron--Frobenius theorem and the assumed irreducibility of $\bm{B}(\bm{a})$ imply the profile $\mathbf{a}^*$ is interior. The fact that $\mathbf{B}(\mathbf{a}^*)$ has $1$ as an eigenvalue implies, as discussed in the proof of Proposition \ref{prop:pareto} above, the existence of a $\bm{\theta} \in \mathbb{R}_{++}^n$ such that $\bm{\theta} \mathbf{B}(\mathbf{a}^*) = \bm{\theta}$. 

Let us normalize utility functions so that $J_{ii}(\mathbf{a}^*) = -1$. We will guess Lindahl prices $$P_{ij} = \theta_i J_{ij}(\mathbf{a}^*) \text{ for $i \neq j$.} $$ For notational convenience, we also define a quantity $P_{ii}=\theta_iJ_{ii}(\mathbf{a})$. To show that at these prices, actions $\mathbf{a}^*$ are a Lindahl outcome, two conditions must hold. The first is the budget-balance condition ($\text{BB}_i(\bm{P})$). Second, agents must be choosing optimal action levels subject to their budget constraints, given the prices.

First, we will show that  at the prices we've guessed, equation $\text{BB}_i(\bm{P})$ holds with equality and so each agent is exhausting his budget:
\be   \sum_{j : j\neq i} P_{ij}a^*_j - a^*_i   \sum_{j : j \neq i} P_{ji} =0.\label{exhausting}\ee To this end, note that $\bm{\theta} \mathbf{B}(\mathbf{a}^*) = \bm{\theta}$ is equivalent to
$$   \sum_{i \in N} \theta_i J_{ij}(\mathbf{a}^*)=0 \quad \quad \Leftrightarrow \quad \quad   \sum_{j \colon\ j \neq i} P_{ij}=-P_{ii}, \label{sum-prices}$$
where the rewriting on the right is from our definition of the $P_{ij}$.  Now, the equation
(\ref{exhausting})  that we would like to establish becomes $$ \sum_{j\colon j\neq i} P_{ij}a^*_j + a^*_iP_{ii}=0$$ or $\mathbf{P} \mathbf{a}^*=\mathbf{0}$. Because row $i$ of $\mathbf{P}$ is a scaling of row $i$ of $\mathbf{J}(\mathbf{a}^*)$, this is equivalent to $\mathbf{J}(\mathbf{a}^*) \mathbf{a}^*  = \bm{0}$. But this is readily seen to be equivalent $\mathbf{B}(\mathbf{a}^*) \mathbf{a}^*=\mathbf{a}^*$, which we assumed.

It remains only to see that each agent is optimizing at prices $\mathbf{P}$. The essential reason for this is that  price ratios are equal to marginal rates of substitution by construction. Indeed, when all the denominators involved are nonzero, we may write:
\be \frac{P_{ij}}{P_{ik}} = \frac{\theta_i J_{ij}(\mathbf{a}^*)}{\theta_i  J_{ik}(\mathbf{a}^*) } = \frac{J_{ij}(\mathbf{a}^*)}{J_{ik}(\mathbf{a}^*)}. \label{eqn:PJratio}\ee Since $P_{ii}$ is minus the income that agent $i$ receives per unit of action, this checks that each agent is making an optimal effort-supply decision, in addition to trading off all other goods optimally.

The converse implication---that if $\mathbf{a}^*$ is a nonzero Lindahl outcome, then $\mathbf{B}(\mathbf{a}^*) \mathbf{a}^*  = \mathbf{a}^*$---is much easier. A nonzero Lindahl outcome $\mathbf{a}^*$ can be shown to be interior. Given this, and that agents are optimizing given prices, we have $ {P_{ij}}/{P_{ik}} = {J_{ij}(\mathbf{a}^*)}/{J_{ik}(\mathbf{a}^*)}, $ which echoes (\ref{eqn:PJratio}) above. In other words, each row of $\mathbf{P}$ is a scaling of the same row of $\mathbf{J}(\mathbf{a}^*)$. Therefore, the condition that each agent is exhausting his budget,\footnote{This follows because each agent is optimizing given prices, and by the assumed irreducibility of $\mathbf{B}(\mathbf{a})$, there is always some contribution each agent demands.} which can be succinctly written as $\mathbf{P} \mathbf{a}^* = \mathbf{0}$, implies that $\mathbf{J}(\mathbf{a}^*)\mathbf{a}^*=\mathbf{0}$, i.e., $\mathbf{B}(\mathbf{a}^*) \mathbf{a}^*=\mathbf{a}^*$.

\subsection{Explicit Formulas for Lindahl Outcomes }\label{sec:explicit}
The eigenvalue and eigenvector conditions we have derived are implicit in that $\mathbf{a}$ appears as an argument of $\mathbf{B}$ in both Proposition \ref{prop:pareto} and Theorem \ref{thm:main}. However, under specific functional forms, we can obtain closed-form characterizations of Lindahl actions in terms of centralities in a fixed network.

The preferences we consider are:
\be u_i(\mathbf{a}) = - a_i + \sum_{j} \left[\alpha G_{ij} a_j + H_{ij} \log a_j \right],  \label{eqn:family} \ee
where $\mathbf{G}$ and  $\mathbf{H}$ are nonnegative matrices (networks) with zeros on the diagonal (no self-links) and $\alpha < 1/r(\mathbf{G})$. Let $h_i=  \sum_{j} H_{ij}$. For any preferences in this family, the centrality property ($\mathbf{a} =  \B(\mathbf{a})\mathbf{a}$) reduces to the equation $\mathbf{a}= \mathbf{h}+\alpha\mathbf{G}\mathbf{a}$.

If $\alpha=0$, then $a_i=h_i$ and $i$'s Lindahl action is equal to the number of $i$'s neighbors in $\mathbf{H}$, which is $i$'s \emph{degree centrality}. If, instead, $h_i=1$ for all $i$, then $\mathbf{a}=\left[\mathbf{I} -\alpha \mathbf{G}\right]^{-1}\mathbf{1}$. This is the vector of \emph{Bonacich centralities}, which are weighted sums of the numbers of walks of various lengths.\footnote{For background, see \citet*[Section 3]{BallesterCalvoZenou} and \cite[Section 2.2.4]{JacksonBook}.} Finally, as $\alpha$ approaches $1$, agents' actions become proportional to their normalized eigenvector centralities in $\mathbf{G}$. 

This specification gives explicit centrality expressions for the Lindahl actions, paralleling a characterization of Nash equilibrium actions in a different setting \citep{BallesterCalvoZenou}.

\section{Concluding Discussion}

The interdependence or ``networked'' nature of economies is one of their central features. Related prior work in networks has mainly focused on public goods games where agents unilaterally decide how much to contribute in a Nash equilibrium. A rich theory on the interplay between network features and economic outcomes has emerged in that case.\footnote{See, e.g., \citet{BallesterCalvoZenou}, \citet{bramoulle-kranton-damours}, and \citet{ggg}.} There, the key network is one that represents the strategic complements or substitutes: how one agent's actions affect others' marginal returns to contributing; these strategic interactions are captured in cross partials of utility. In the present theory, the network we construct simply tracks marginal externalities---first derivatives.

Perhaps the closest prior work \citet{ghosh2008charity}, a model of negotiations in a linear environment. Agents benefit linearly from their neighbors' contributions, with a cap on how much each can contribute. Their main result is that there is an equilibrium of their game that achieves the maximum feasible contributions if and only if the largest eigenvalue of a network weight matrix is greater than one---a result generalized by our Proposition \ref{prop:pareto}. \citet*{du2015competitive} study a competitive exchange economy with particular parametric (Cobb-Douglas) preferences, and characterize its price equilibria in terms of a matrix describing the preferences, which in some ways resembles our Theorem \ref{thm:main}.

The reliance of all these prior results on parametric assumptions leaves open the possibility that they are dependent on the functional forms. Our contribution is to show, without parametric assumptions,  that outcomes of negotiations with externalities can be characterized exactly by eigenvalue or centrality properties. In doing this, we give a new economic angle on these important mathematical concepts. In the opposite direction, the connection offers a new way to approach economic questions with results on positive matrices. Sections \ref{sec:essential} and \ref{sec:subdividing} give some first ideas on how to use the connection, and raise many questions that we hope the techniques will help with.

\begin{tiny}
\bibliographystyle{ecta}
\bibliography{shortbib}
\end{tiny}

\end{document}